# On stabilizing the variance of dynamic functional brain connectivity time series.

**William Hedley Thompson**  WILLIAM.THOMPSON@KI.SE
*Department of Clinical Neuroscience*
*Karolinska Institutet,*
*Stockholm, Sweden*

**Peter Fransson**  PETER.FRANSSON@KI.SE
*Department of Clinical Neuroscience*
*Karolinska Institutet,*
*Stockholm, Sweden*

**Abstract**

Assessment of dynamic functional brain connectivity (dFC) based on fMRI data is an increasingly popular strategy to investigate temporal dynamics of the brain's large-scale network architecture. Current practice when deriving connectivity estimates over time is to use the Fisher transform which aims to stabilize the variance of correlation values that fluctuate around varying true correlation values. It is however unclear how well the stabilization of signal variance performed by the Fisher transform works for each connectivity time series, when the true correlation is assumed to be fluctuating. This is of importance because many subsequent analyses either assume or perform better when the time series have stable variance or adheres to an approximate Gaussian distribution. In this paper, using simulations and analysis of resting-state fMRI data, we analyze the effect of applying different variance stabilization strategies on connectivity time-series. We here focus our investigation on the Fisher transform, the Box Cox transform and an approach that combines both transforms. Our results show that, if the intention of stabilizing the variance is to use metrics on the time series where stable variance or a Gaussian distribution is desired (e.g. clustering), the Fisher transform is not optimal and may even skew connectivity time series away from being Gaussian. Further, we show that the suboptimal performance of the Fisher transform can be substantially improved by including an additional Box-Cox transformation after the dFC time series has been Fisher transformed.

**Keywords: dynamic functional connectivity, fMRI, time series, Box Cox transformation, Fisher transform, variance.**

# 1 Introduction

The idea behind dynamic functional connectivity (dFC) in fMRI is both simple and appealing. It builds upon the previous successes of static functional connectivity (sFC) in which the covariance between two brain regions is quantified and subsequently an inference is made regarding their relationship that is estimated over a fixed (static) time interval. To date, sFC has been applied to many areas of the brain research, from a general understanding of the network topology of the brain (Greicius et al., 2003; De Luca et al., 2005; Fransson 2005), task modulation (Fransson 2006), neurodevelopment (Power et al., 2010) and clinical applications (Fox & Greicius 2010). In the case of dFC, quantitative studies of the fluctuations of signal covariance over time offers a possibility to explore the dynamics of the brain and it has already found applications; from understanding basic brain processes such as levels of consciousness (Bartffeld et al 2015), mind-wandering (Schafer et al 2014) and development (Hutichison & Morton 2015), to clinical applications such as depression (Kaiser et al 2015) and schizophrenia (Damaraju et al 2014; Ma et al 2014).

dFC offers an exciting perspective, but it is not without controversy. A substantial amount of work has been devoted to the issue of the accuracy of the dynamic functional connectivity estimates obtained by the sliding window method (Allen et al., 2013; Hutchison et al., 2013; Leonardi & De Ville 2015; Hindrinks et al., 2015; Zalesky and Breakspear, 2015) and whether the fluctuations in BOLD connectivity accurately reflect the presumed underlying neuronal dynamics (Tagliazucchi et al 2012, Chang et al 2013; Magri et al 2012; Thompson et al 2013; Thompson et al 2014; see also Keilholz 2014). That being said, in this paper we assume that sliding window estimates of dFC are indeed fluctuations that reflect neuronal activity. The next question that arises is how to quantify fluctuations in dFC time series in an accurate manner. In this article we want to highlight a methodological aspect of dFC analysis which often is taken for granted in the data pre-processing pipeline, namely the how to stabilize the variance of time-series of co-variance estimates.

A dynamic functional brain connectivity time series is usually created by estimating the connectivity (most often Pearson correlation coefficients) for multiple time-points across two nodes (often voxels or regions of interests) for which BOLD fMRI signal intensity time series have been extracted. The resulting time series represents the degree of connectivity that fluctuates as a function of time between the two nodes. Importantly, a correlation coefficient is bound to range between -1 and 1 and its expected variance is smaller as the correlation coefficient increases. The aim of stabilizing the variance is to disassociate the variance from its mean. For example, if the sampled correlations from subjects in group A are centered around a true correlation value of $\rho=0.6$, and the sampled correlations from subjects in group B are centered around a true correlation value of $\rho=0.2$, one would expect the variance from group B to be larger than the variance in group A. Hence, the purpose of stabilizing the variance is to achieve estimates of variance that is unbiased from the magnitude of the true correlation. In this outlined example, the Fisher transform performs well in terms of stabilizing the variance between groups.

However, in the case of dFC fMRI we are met with a number of circumstances that complicate matters in terms of stabilizing the variance compared to the previous example. Given the assumption that dFC reflects neuronal fluctuations in brain connectivity, (i.) the dFC time series reflects fluctuating true connectivity values; (ii.) estimates of connectivity for neighboring time-points are not independent of each other when using the sliding window method; (iii.) given (i), the variance will then fluctuate as the true correlation value, and it is unknown how long the true correlation dwells with different true correlation values. Together these properties differ from the original purpose of the Fisher transformation. dFC requires stable variance across an entire time series of fluctuating values whereas the Fisher transform is intended for stable variance around estimates of a true correlation value, making the variance independent of their correlation magnitudes.

In the neuroimaging dFC literature, the variance is often stabilized by applying the Fisher transform to the connectivity time series. Generally, there are good reasons for doing so, since a reasonably stable signal variance is required to be able to accurately quantify changes in dynamic brain functional connectivity, which often is the primary goal of the analysis. In this study we test how the Fisher transform, the Box Cox transform and a combination of both transforms performs in terms of achieving an estimate of co-variance that is unbiased by the mean correlation estimate. Further, we quantify to which extent the within-time series variance is stable for the tested transformations when applied to dynamic functional connectivity fMRI time series and quantify how Gaussian the resulting distributions are. This was done because both transformations often create approximate Gaussian distributions of the data, which has a stable variance.

## 2   Method

### 2.1   Simulations

All simulations were done in Matlab (2014a) using the "normrnd()" and "rand()" functions to create random univariate and Gaussian distributions with different means and variances. A Fisher transform was then applied and different properties were then quantified. The procedure for the simulations is outlined in the results section.

### 2.2   Data and preprocessing steps

One resting-state fMRI session (6 minutes long; 3 Tesla, TR = 2000 ms, TE = 30 ms) from 48 healthy subjects was used in the analysis (19-31 years, 24 female). Two subjects were excluded from the analysis due to incomplete data. The fMRI data was downloaded from an online repository: the Beijing eyes open/eyes closed dataset available at http://fcon_1000.projects.nitrc.org/indi/IndiPro.html (Liu et al. 2013). Each functional volume comprised 33 axial slices (thickness / gap= 3.5 / 0.7 mm, in-plane resolution = 64 × 64, FOV = 200 × 200 mm). The fMRI dataset from each individual contained 3 different resting-state sessions, each 6 minutes long (two eyes closed sessions and one eyes open session). We only used data from the eyes-open condition, which was recorded in either in the $2^{nd}$ or $3^{rd}$ session (counterbalanced order with respect to the second eye-closed session). Further details regarding the scanning procedure are given in Liu et al., 2013.

fMRI data was pre-processed using Matlab (Version 2014b, Mathworks, Inc.), the CONN (Whitfield-Gabrieli & Nieto-Castanon 2012) and SPM8 (Friston et al., 1995) Matlab toolboxes. Resting-state fMRI data was realigned and then normalized to the EPI MNI template as implemented in SPM. Spatially smoothing was then applied using a Gaussian filter kernel (FWHM = 8 mm). Additional image artifact regressors attributed to head movement (Van Dijk et al., 2012; Power et al., 2012) were derived by using the ART toolbox for scrubbing

(www.nitrc.org/projects/artifact_detect/). Signal contributions from white brain matter, cerebrospinal fluid and head-movement (6 parameters), and the ART micro-movement regressors for scrubbing, were regressed out from the data using the CompCor algorithm (Behzado et al 2007), as implemented in CONN. After regression, the data was band-passed between 0.008-0.1 Hz, as well as linearly detrended and despiked.

264 regions of interest (ROIs, sphere with a 5mm radius) were placed throughout the brain according to the parcellation scheme provided in Power et al 2011. All analysis, except for the mean-variance plot in figure 5F, was restricted to the default mode network (DMN) (motivation for this is given below). 58 of the ROIs were assigned to be DMN, resulting in 1653 edges per subject.

Static functional connectivity (sFC) was computed by the Pearson correlation coefficient across all time-points.

### 2.3 Estimation of dynamic functional connectivity.

We used the sliding window method to create dFC time series. For each time-point, $t$, a Pearson correlation coefficient was estimated using +/- 22 time-points (45 volumes in total equaling 90 seconds) for each combination of ROIs (nodes). This resulted a unique connectivity time series with a length of 196 time-points (reduced from the full 240 time-points due to the window size), for each subject and condition, with 264x264 correlation values at each time-point. The chosen length of the sliding time-window is well in line with "rules of thumb" that have been previously suggested for sliding window analysis (Leonardi and Van De Ville 2015, Zalesky and Breakspear, 2015). Different transformations (see below) were then applied to the dFC time series data.

### 2.4 Transforms used to stabilize the variance of dFC time series.

We wanted to compare the statistical properties of four different distributions of dFC time series. First, we considered the "raw" dFC time series, calculated by the sliding-window technique. Second, we considered the Fisher transform applied to the dFC time series. Third, we analyzed the Box-Cox transformed dFC time series. Finally, we investigated a combined approach where the dFC values were first Fisher transformed and subsequently Box-Cox transformed.

#### 2.4.1 Fisher transformation

The Fisher transform takes the bounded distribution of correlation coefficients ($r$) and makes it unbounded so that the variance is independent of the magnitude of the correlation coefficient by:

$$z = \frac{1}{2} \ln\left(\frac{1+r}{1-r}\right)$$

The estimates of z generally approximate to a Gaussian distribution.

#### 2.4.2 Box-Cox transformation

An alternative way to stabilize the variance of a non-Gaussian distributed dataset is the Box-Cox transformation (Box & Cox 1964). The Box-Cox transformation, which is a power transformation, is given by

$$y_i^\lambda = \begin{cases} \dfrac{y_i^\lambda - 1}{\lambda} & \text{if } \lambda \neq 0 \\ \ln(y_i) & \text{if } \lambda = 0 \end{cases}$$

in which $\lambda$ is set by the estimated maximum likelihood for each edge, with $\lambda$ ranging from -5 to 5 in increments of 0.01. $y$ is the time series that it is applied to and $i$ is the index of the time series.

The distribution of the $\lambda$ values for within-DMN connectivity edges (total number of edges = 1653) is shown in Figure 1. The Box-Cox transformation was applied to both the raw dFC time series (Figure 1A) and on the Fisher transformed dFC time series (Figure 1B).

Of note, the Box-Cox transformation cannot handle values less than zero due to the natural logarithm when $\lambda = 0$. To deal with this problem in this study, the smallest value of each dFC time series was scaled to 1. This step was mandated by the fact that power transformations can perform differently between 0 and 1, and by scaling the smallest value to 1, we ensured that a similar type of power transformation was applied to all edges. The Box-Cox transformed dFC time series was subsequently scaled back so that the post-Box Cox mean equaled the mean of the raw dFC time series. In the combined case when the Box-Cox transformation was applied to the data after a Fisher transformation, the mean was scaled to the mean of the Fisher transformed dFC time series.

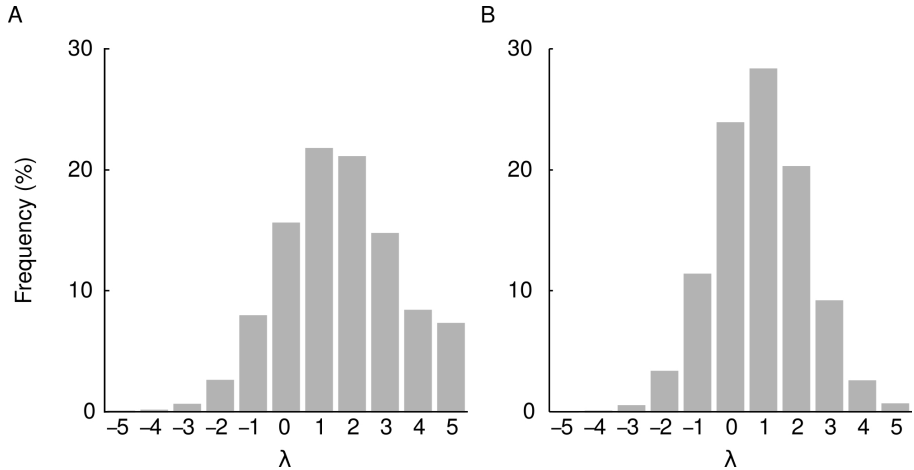

*Figure 1: **T**he distribution of the best-fitting value of the parameter $\lambda$ (range -5 to 5) included in the Box-Cox transformation when applied to all edges within the default mode network (pooled across 46 subjects). (A) $\lambda$ values obtained when applying the Box-Cox transformation on the raw dFC values. (B) $\lambda$ values obtained from the Box-Cox transformation on dFC values that first have been subjected to the Fisher transform.*

## 2.5  Quantifying variance stability

To quantify the within-time series variance stability we performed a median split on the dFC time series, dividing the time series data into two separate partitions of connectivity estimates. Hence, the absolute difference of the variance in the upper and lower median partition was calculated, providing an estimate of the stability of the variance. The median split allows for an estimate of the stability in the variance such that a smaller absolute difference between the divisions (i.e. approaching 0) entails a stable variance. To make the variances of the different transformations comparable, each dFC time series was scaled between 0 and 1 before performing the median split.

## 2.6  Quantifying the Gaussian distributions

A time series that adheres to a Gaussian distribution will by definition have stable variance. We were therefore interested to assess how Gaussian each time series was after each transformation. We used two different methods for this: the Shapiro-Wilk (SW) test statistic and skewness (s).

### 2.6.1  Shapiro-Wilk statistic

In order to evaluate how closely each of the four distributions of raw, Fischer, Box-Car, Fischer and Box-Car transformed dFC data followed a Gaussian distribution, we used the Shapiro-Wilk test, which is one of the most robust methods to estimate the normality of a given distribution (Razali & Wah 2011), in particular for the case when the number of values in each distribution to be tested is rather small.

The Shapiro-Wilk test statistic, W, is calculated by

$$W = \frac{\left(\sum_{i=1}^{n} a_i x_i^2\right)}{\sum_{i=1}^{n} (x - \bar{x})}$$

where $x$ is the ordered statistic and $\bar{x}$ is the mean. The constants $a_i$ are given by

$$(a_1, a_2, \ldots, a_n) = \frac{m^T V^{-1}}{\left(m^T V^{-1} V^{-1} m\right)^{\frac{1}{2}}}$$

where $V$ is the covariance of the ordered statistic, and $m$ is the expected value of the ordered statistic, given a Gaussian distribution (Shaprio & Wilk 1965). The test statistic $W$ was then normalized according to (Royston 1992, 1993)

$$z = \frac{(\log(1 - W) - \mu)}{\sigma}$$

where μ and σ are the mean and variance of the expected Gaussian distribution. The test statistic is usually compared against a chosen p-value threshold. We compared the SW-statistic between the proposed variance-stabilizing transforms, where a smaller normalized SW-statistic suggests a better fit to a Gaussian distribution.

### 2.6.2 Skewness

An additional way to quantify the Gaussian distributions is to examine its skewness (*s*) which is defined as:

$$s = \frac{\left(\frac{1}{n}\sum_{i=1}^{n}(x_i - \bar{x})^3\right)}{\left(\sqrt{\frac{1}{n}\sum_{i=1}^{n}(x_i - \bar{x})^2}\right)^3}$$

for which $x$ is the distribution to be tested (i.e. the dFC time series), $s$ the skewness of the distribution and $\bar{x}$ is the mean of the distribution. In essence, the skewness is a measure of how much a given distribution is shifted away from a Gaussian distribution. A large skewness entails a more fat-tailed distribution. For a unimodal distribution, a positive skewness entails that there will be a fat-tail to the right of the mean, a negative skewness means a fat-tail to the left of the mean. A skewness of 0 would imply a more Gaussian distribution.

### 2.7 Data visualization and parameter choices

As stated above we limited our analysis to the subset of positive edges between all of the 58 nodes within the default mode network (DMN). The reason for our choice of restricting our analysis to within-DMN brain connectivity in the present analysis is that this subset of brain nodes generally shows a strong degree of connectivity and that a large number of edges within the DMN have correlation coefficient values that span the range between 0 and 1 (see also Supplementary Figure 1), which is the range we chose to visualize many of the results. Our motivation for this is that if, on the other hand, we had included edges between all 264 nodes, this would have resulted in that a considerably higher number of edges with lower connectivity to be included in the results. The results displayed between 0 and 1 are binned and averaged by the sFC in 0.025 increments. Notably, the sFC connectivity value, which are displayed on the x-axis, reflect the mean connectivity value for each of the different transformations (Supplementary Figure 1). Although the inclusion of all edges would quantitatively change the results, it is important to point out that our main results and conclusions regarding how to stabilize the variance for dFC time series would still be valid. So, in the interest of clarity and visibility for the figures that present the results, we limited our presentation to be based on all edges within the DMN. The one exception to focusing on DMN edges was the mean-variance plot shown in Figure 5F which is based on edges using all 264 ROIs.

# 3    Results

## 3.1    Simulations that highlight the potential problem with the Fisher transformation when applied to dFC time series that fluctuates around a non-zero mean.

To illustrate the complications that might be inflicted by the Fisher transform when applied to dFC time series, let us start by giving an example of how the Fisher transform behaves when applied to a univariate random distribution of connectivity estimates. We simulated a univariate distribution bounded to the range -0.95 and 0.95 (Figure 2A). If we apply the Fisher transform to the distribution shown in Figure 2A, it will transform it to be both unbounded from the -1 to 1 range and pushing it towards a Gaussian distribution (Figure 2B).

Thus, every dFC time series will have its unique distribution of connectivity values, with a different variance and mean. To exemplify the effect of the Fisher transform on dFC time series, let us first consider a simulated and normalized distribution of a single dFC time series (Figure 2C and Figure 2D). Obviously, raw dFC correlation values will not be a Gaussian distribution (otherwise there would be little point in using variance stabilizing transformations), but it still helps illustrate that post-Fisher transformed time series can have neither stable variance nor adhere to a Gaussian distribution. In our example, the Fisher transform will then skewed the distribution slightly in the positive direction (direction of arrow in Figure 2D). To better visualize this potential skewing effect, we simulated dFC time series with 500 time points, incrementally shifting the mean ($\mu$) of the time series ($\mu$ ranged from -0.9 to 0.9, shifted in 0.01 increments; variance kept equal at $\sigma^2=0.025$). By applying the Fisher transform to the complete set of simulated dFC time series we observed that, as the mean of the dFC time series is situated further away from zero, the skewness increases (Figure 2E). We also note that the difference in signal variance between the raw and the Fisher transformed dFC time series increases as a function of the difference with respect to the zero-mean distribution (Figure 2F). It is noteworthy that both the observed difference in variance and the difference in skewness correlate with each other (Figure 2G). The correlation shown in Figure 2G illustrates the fact that biases in dFC time series variance might be introduced as an effect of using the Fisher transform and causes the variance of the dFC to destabilize rather than the intended outcome of stabilizing the variance.

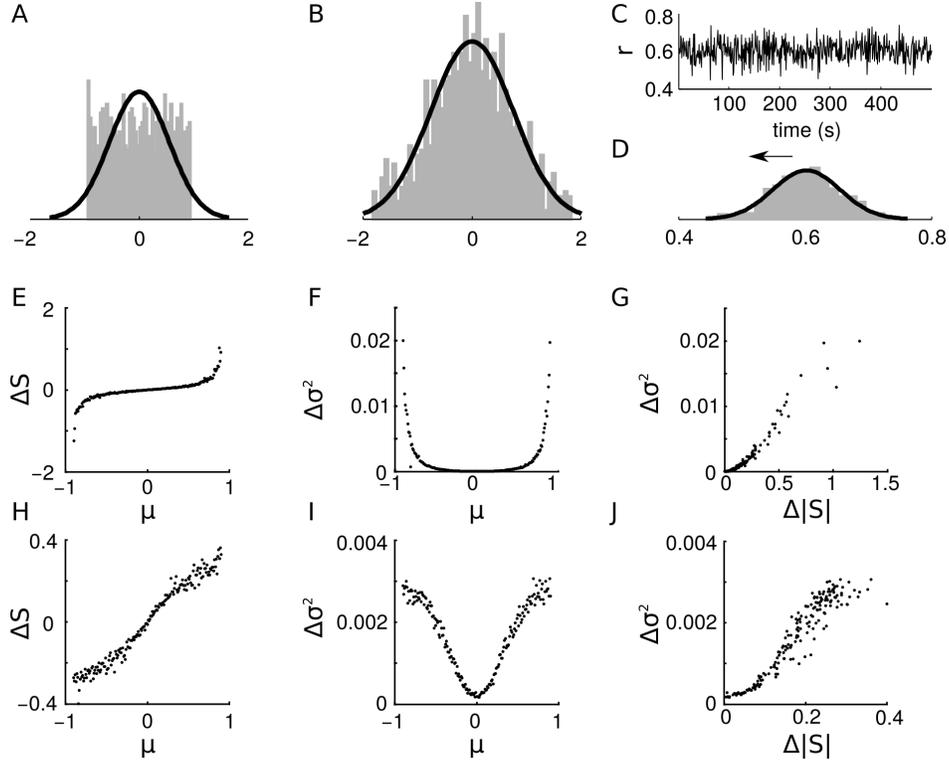

*Figure 2:* Results from simulations that serve to illustrate the potential difficulties that are involved when applying the Fisher transform to dFC time series. Panel (A) shows a simulated univariate distribution of correlation values that ranges between -0.95 and 0.95. Panel (B) shows the corresponding Fisher transformed data. (C) A simulated dFC time series where the connectivity estimates follow a Gaussian Distribution. Panel (D) shows the Gaussian distribution of the data shown in panel C. The arrow indicates the shift that will occur after a Fisher transform, which will create a fat-tail to the right. Panel (E) shows the relationship between the mean ($\mu$) and skewness (S) for 500 simulated dFC time-courses after Fisher transformation ($\mu$ ranged from 0.9 to 0.9 in 0.01 increments, variance kept equal at $\sigma^2=0.025$). The difference in signal variance ($\Delta\sigma^2$) as a function of mean ($\mu$) after Fisher transformation is shown in panel (F) and panel (G) shows the correlation between the observed difference in variance and skewness after Fisher transformation. Panels (H-J) shows the corresponding plots for the case when the variance was scaled negatively with respect to the mean ($\sigma^2=(1-|\mu|)/10$) before the Fisher transformation.

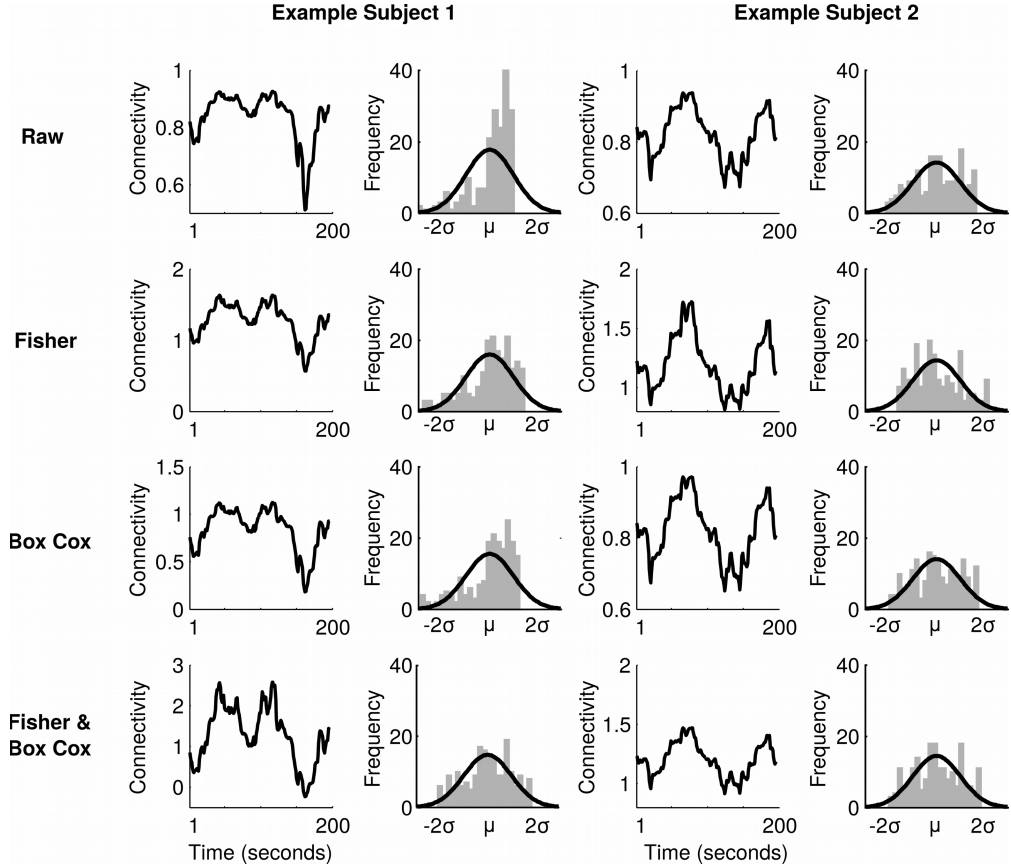

*Figure 3: Dynamic functional connectivity time series and their corresponding distributions computed for a single edge in the default mode network in two different subjects. The top row shows the raw dFC time series and their corresponding distributions together with a fitted Gaussian distribution (solid line). The second row shows the dynamic connectivity values after Fisher transformation. The third row display the case of applying the Box-Cox transformation. The last row shows the result from applying first the Fisher and then the Box-Cox transformation on the same data.*

Since we know that the variance of the raw dFC time series decreases when its mean is shifted away from zero due to the boundedness of the correlation coefficient (e.g. Thompson & Fransson 2015), it would be of interest to investigate what happens in terms of variance stability after the Fisher transformation in a scenario which more accurately reflect the mean-variance relationship obtained for recorded resting-state fMRI data. Accordingly, we repeated the previous simulation but scaled the variance negatively as a function of increases in the mean of the Gaussian distribution ($\sigma^2=(1-|\mu|)/10$). The results shown in Figures 2H (skewness), 2I (difference in variance) and 2J (correlation between skewness and variance) suggests that also in the case of a negatively scaling between variance and the mean, the Fisher transformation has similar variance destabilizing effects.

So far, our simulations have shown the potential of dFC time series being skewed away from a Gaussian distribution (and hence having unstable variance) when using the Fisher transform.

ON STABILIZING THE VARIANCE OF dFC TIME SERIESWe have also observed that the degree of skewness is proportional to the mean connectivity value in the dFC time series. Although our simulated dFC time series are Gaussian, this will not be the case for recorded resting-state fMRI data connectivity estimates. We now turn our attention to investigating how recorded dFC fMRI time series are distributed and study how the different transforms are able to stabilize the variance.

**3.2 A qualitative assessment of the effect of the Fisher, Box-Cox transform and their combined use on empirical dFC resting-state fMRI data.**

To illustrate the effects of the different transformations on empirical dFC time series data we start by presenting data from two subjects and a single edge between two nodes located in the DMN. Figure 3 shows the time series of dFC correlation coefficients and their corresponding distributions for four different cases. If we start by assessing the results obtained for the first subject, we note that the distribution of the raw dFC correlation coefficient values has a sharp peak of correlation coefficient values that are larger than its mean. This peak is accompanied by a greater spread of correlation coefficients in the lower regime. If we examine the corresponding Fisher transformed distribution of dFC values (Figure 3, second row), we can in this case observe that although the Fisher transform alters the distribution so that it becomes unbounded by the -1 to 1 restriction, it still performs poorly in terms of creating a Gaussian distribution of the dFC correlation values. Similarly, the Box-Cox transformation used in isolation (Figure 3, third row) fares no better. However, the combined approach provides a rather good approximation to a Gaussian distribution (Figure 3, last row).

Next, we proceed by examining the performance of the suggested transforms on the data obtained from the second subject (shown in the right column in Figure 3). In this case, another potential problem with the Fisher transformation becomes apparent. Here, the raw dFC time series constitutes a distribution that is fairly Gaussian. When we apply the Fisher transform, we can observe that the distribution is skewed away from a Gaussian distribution. However, both the Box-Cox and the case of a combined transformation create distributions of dFC correlation coefficients that qualitatively approximate a Gaussian distribution rather well.

From the examples shown here, where data was taken from a single edge between two nodes, in the DMN for two different subjects, we have illustrated two possible obstacles that pertains of the Fisher transformation in the context of stabilizing the variance of a dFC time series. First, the Fisher transform may fail to stabilize the variance and thereby provide a poor Gaussian distribution approximation for the data, if the raw dFC time series is heavily skewed as shown for the first subject in Figure 3. Second, as exemplified for the second subject shown in Figure 3, it may skew the distribution in the opposite direction, again performing poorly in terms of achieving a Gaussian distribution. We also note, although at present only qualitatively in Figure 3, that applying the Box-Cox transformation subsequent to the Fisher transform seems to achieve the best performance for both subjects. This is perhaps not overly surprising, since the dFC time series become unbounded after the applying the Fisher transform and despite the skewness that might occur due to the Fisher transform, the Box-Cox manages to correct for the skewness. Perhaps due to the boundedness of the raw dFC correlation values, that the Box-Cox transformation alone does not adequately correct for this and might actually perform worse than the combined transform approach.

An important observation from the results shown in Figure 3 is that a discussion of the stability of the variance from the tested transformations in terms of obtaining Gaussian distributions is warranted. This is because the data does not have stable variance and the data could potentially adhere to another type of distribution (e.g. a multimodal distribution). Thus, when we quantify the SW-statistic and skewness of dFC time series to quantify the extent of approximating a Gaussian distribution, this issue is related to the stability of the time series variance.

### 3.3    Quantifying the stability of the variance of each dFC time series.

So far, our simulations and examples based on dFC time series from single edges suggest that the Fisher transform may not stabilize the variance for the time-series. We now proceed by showing that the destabilizing effects of the Fischer transform is generally present and that alternative data transformation might perform substantially better in terms of stabilizing dFC time-series variance. We start by examining the difference in variance in terms of the median split approach followed by a quantification of how well the dFC variance follows a Gaussian distribution.

We aimed to quantify the stability of the variance in each time series across all edges and subjects and for all cases of data transforms. Ideally, the variance should be similar throughout the entire length of the dFC correlation value time-series. By splitting the data into two partitions through a median split, we calculated the absolute difference in the variance between partitions. The median split procedure allowed us to investigate if the variance is similar in different parts of the time series. If the distribution is indeed Gaussian, the absolute difference will be 0. Figure 4 shows that the average difference in variance, binned with respect to their static connectivity values, is lowest for the combined Fischer/Box-Cox approach, followed by the Box Cox transform, although it becomes less stable when the amplitude of the connectivity values increases. Larger distances were observed for the Fischer transform and the largest distance in variance was seen for the raw dFC time-series. The median split analysis suggests that the within-time dFC series variance is most stable when the combined Fischer/Box-Cox data transformation strategy is used.

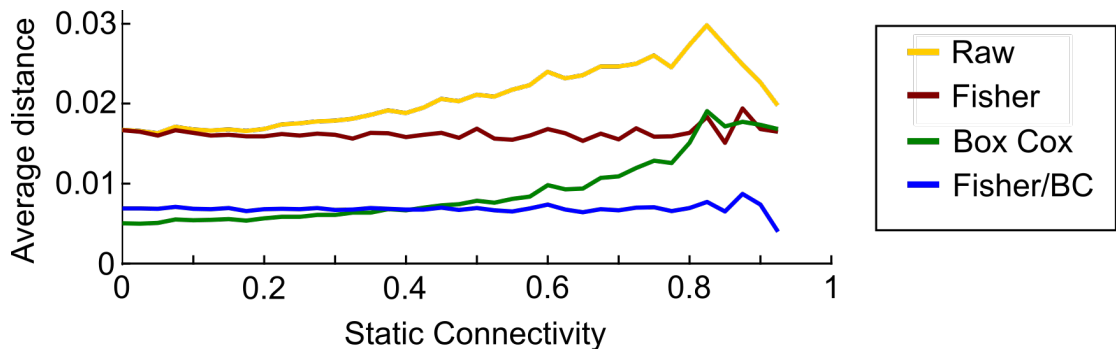

***Figure 4***: *Differences in dFC time-series variance as a function of static connectivity for the raw data and the three variance stabilizing transforms investigated. The plot shows the averaged absolute distance in variance (pooled across edges and subjects) and binned according to its corresponding static connectivity value. The variance of connectivity was estimated as the absolute average distance of the time-series connectivity values when divided into an upper and lower partition by a median split.*

### 3.4 Quantifying normality and skewness of dFC time series.

We now proceed by quantifying and evaluating the performance of the three suggested approaches to stabilize the variance with regards to achieving a Gaussian distribution for each dFC time series. If we by means of applying a variance stabilizing transform to the data obtain an approximate Gaussian distribution, not only will the variance be stable, but also it will also help us to meet some assumptions of the data that is typically made in dynamic brain connectivity analysis.

First, for all DMN edges, we noted that for many of them, their corresponding distributions were classified as being non-Gaussian by the SW tests (Figure 5A). The combined Fisher and Box-Cox transformation approach had the least amount of edges classed as non-Gaussian for significance thresholds ranging from p<0.01 to p<0.00001. Of note, even with the lowest statistical threshold, 30% of edges were still classified as non-normal and for the higher threshold (p<0.01), 84% of all edges were classified as non-Gaussian. The three other cases fared worse in that the raw dFC time series contained 95% to 60% of non-Gaussian edges. The non-Gaussian edges when applying the Fisher transform ranged from 93 to 52% and the Box-Cox transformed data ranged from 87% to 37% of the edges being classified as non-Gaussian. It should be noted that a significant value obtained with the SW statistic disproves the null-hypothesis which is that the distribution of dFC values are normally distributed. Although we may fail in terms of producing an optimal result in which all edges approximately follows a Gaussian distribution for any of the proposed transformation strategies, we can still conclude that large performance differences exist for the three proposed approaches in terms of stabilizing the variance for dFC time series.

Next, we compared the average SW-statistic value for all edges with regard to their sFC correlation coefficient (Figure 5B). We observe that both the raw and Box-Cox transformed dFC time series were classified to be less Gaussian as the static connectivity value increased. The Fisher transformed dFC data behaves worse that the Box-Cox transformed data for lower static connectivity values but are more on an equal footing for the case when the static connectivity value increase above 0.7 and the Box-Cox transform is unable to perform optimally on the bounded data. For the combined Fischer/Box-Cox approach the SW statistic quantified the distributions as being more Gaussian. This finding was consistent across all connectivity values. The combined Fisher and Box-Cox transformed data had the lowest SW-statistic for 64% of all edges, the Box-Cox transformed data had the lowest SW-statistic for 21% of all edges, followed by the Fisher transformed data with 14%. The raw dFC time series data displayed a Gaussian distribution according to the SW-statistic for less than 1% of all edges (Figure 5C).

The skewness ($s$) of the transformed distributions of dFC correlation estimates together with the raw data gives a similar result across the range of static connectivity (Figure 5D). The raw connectivity time series becomes increasingly skewed as a function of the underlying static connectivity (Figure 5D, upper left panel). The Fisher transformed dFC data (Figure 5D, upper right panel) display a skewness in both positive and negative directions. The positive skewness at higher connectivity values is considerably less for the raw dFC time series and this demonstrate how the Fisher transform can skew too much (similar to the results shown in Figure 3B). For the Box-Cox transformed data, the standard deviation was considerably lower than the raw and Fisher transformed data (Figure 5D, lower left panel), but again became destabilized at the higher connectivity values. The combined transformation once again had the best performance by having a low standard deviation in skewness and was stable across the entire range of mean dFC values (Figure 5D, lower right panel).

### 3.5 Fluctuations of variance in dFC time series.

One byproduct of the data transformations examined here is that the variance of a given dFC time series of correlation coefficients can be affected as well. To exemplify this influence, we correlated the between-subject variance of a single edge dFC time series for the different variance-stabilizing transformations examined (see Figure 5E). While all correlations had a low p-value ($p<0.001$, Bonferroni corrected), it is worth noting that, since this is a correlation of the same time series, the non-parametric Spearman coefficient is quite low (often capturing below 25% of the between-subject variance). While this result is perhaps not overly surprising it shows that, without properly accounted for, it may become hard to interpret what any quantitative difference of dFC variance actually represents when contrasting data between groups or conditions.

### 3.6 The mean-variance relationship of dFC time series after a combined Fisher and Box-Cox transformation.

We have previously argued that if the analysis strategy in studies of dynamic brain connectivity is to find interesting changes of connectivity in time (e.g. to make a binary time series of connectivity that used together with temporal graph theory (Thompson & Fransson 2016)), then one has to take into account the mean-variance relationship that exists for dFC fMRI time series (Thompson & Fransson 2015). Thus, it then becomes relevant to investigate if the previously shown relationship between the mean and the variance of individual edges still holds after the combined Fisher and Box-Cox transformation, as the conceptual argument discussed in our previous study was based on non-stabilized variance. Figure 5E shows the variance plotted against the mean for dFC time series taken from all edges across all 264x264 edges and averaged over subjects for the combined Fisher and Box-Cox transformation (Figure 5F). While the variance is quite stable over edges, the fact that there are more edges when the mean is lower leads to a greater probability that some edges here have higher variance. We observe that our previous observation regarding a relationship between the mean and variance of dFC time series still holds after a Fisher and Box-Cox transformation of the dFC data. This finding implies that different (but reasonable) thresholding strategies (based on taking edges with high mean or variance) will yield different time-points for different edges being marked as candidates of significant/interesting connectivity. This occurs because, despite the variance being relatively stable, the mean-variance dependence still holds and choice of analysis strategy to be used in future studies needs to be carefully considered.

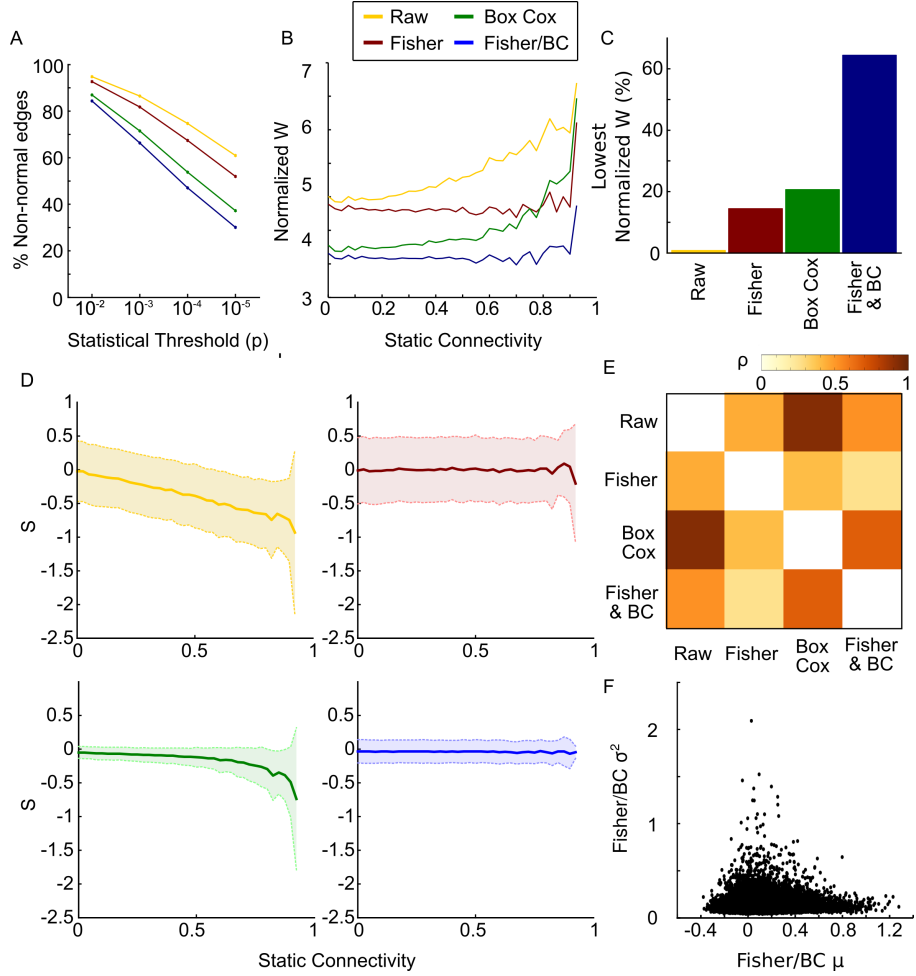

*Figure 5:* *(A) The normalized SW-statistic for all edges within the DMN (pooled across subjects) plotted against their static functional connectivity values for the three different variance-stabilizing transformations together with the raw dFC distribution. Error bars show the standard deviation of the edges in the bin. BC denotes Box-Cox. (B) average SW-statistic across all within-default mode network edges. Error bars shows the standard deviation of the averaged SW-statistic (W) between subjects. (C) The percent (of edges per subject) for each of the four dFC time series that had the lowest SW-statistic. (D) Skewness for the four different distributions (colors correspond with panels A and B) along different static functional connectivity values for the four different transformation possibilities. Error bars show standard deviation. (E) Spearman rank coefficient of the between-subject variance for the four different transforms computed for a single edge. All values are significant at p<0.001, Bonferroni corrected. Note that the Spearman rank coefficient is quite low considering these are correlations of the variance related to the same time series (bar differences caused by the transformations). (F) Mean-variance relationship of the Fischer and Box-Cox transformed connectivity time series. Each dot represents the average mean and the average variance for each edge across all subjects*

## 4 Discussion

From the results presented in this work, we can draw three conclusions. First, dFC time series in terms of within-time series variance stability and degree of skewness compared to a Gaussian distribution, is improved when using the combined Fisher/Box-Cox transformation approach compared to using the prevailing Fisher transformation strategy. Second, the problem of the mean-variance relationship for dFC data (Thompson & Fransson 2015) is still present in data even after applying the combined Fisher/Box-Cox transform, meaning that the process of single out time-points of interest in different connectivity time series will vary depending on chosen thresholding strategy. Third, each of suggested variance-stabilizing transformations display a relatively low degree of correlation across subjects when considering that the correlation are applied to the dFC variance of the exactly same time series with the exception of different transforms being applied (Figure 5F). This is most likely due to a non-linear effect on the dFC variance introduced by the variance-stabilizing transforms. Specifically, the non-linear effect originates in the fact that when the investigated variance-stabilizing transforms are applied to dFC time series, their effect is not uniform across the range static connectivity correlations. Thus, the task of quantifying the variance of dFC time series may pose a difficult problem, since the non-linear effects imposed by the Fisher and/or Box-Cox transforms may artificially inflate or deflate the true variance of a dFC time series. And, to our knowledge, it is uncertain at this time how this can be accounted for since the true variance of a given dFC time series is unknown. This means that attention has to be paid to this dependency when contrasting the variance of dFC time series between different task conditions.

One might ask oneself what is the purpose of stabilizing of the variance? There are two good reasons why one may wish to do this. (i.) to achieve a better estimation of whatever metric of dynamic functional connectivity that is being estimated (ii.) to allow for parametric statistic testing. While the second possibility is feasible using variance-stabilizing techniques, there is no reason (apart from perhaps simplicity) to abandon previously proposed non-parametric testing procedures that has been proposed for dFC studies (Hindrinks et al 2015). Critically, the first possibility can offer more accurate and better performing analysis steps. Various data processing steps that are commonly applied in studies of dynamic functional brain connectivity, for example clustering, will often benefit in performance by transforming the data so that its features adhere to a Gaussian distribution. We stress that the value of the proposed pipeline and our motivation for this study was primarily to achieve an increase in performance in terms of estimating dynamic fluctuations in brain activity, not to replace previous proposals of non-parametric statistics.

It is perhaps pertinent to ask the question if the problem of variance stabilization in the context of dFC time series may after all not pose such a large methodological difficulty as it has been portrayed in this paper? To begin to provide an answer, one may ask oneself when could it be problematic to use the Fisher transformation with the intent to stabilize the variance for dynamic functional connectivity time series? We identify two possible problems with using the Fischer transformation on dFC time series. First, it becomes harder to interpret the variance measures of the time series when there are shifts in dFC variance imposed by the Fisher transform and some of edges are skewed away from normality by the transformation. Second, a substantial part of the dynamic functional brain connectivity literature uses k-means clustering techniques on Fisher transformed data for which the stabilized variances will often dramatically improve clustering performance. dFC studies considering these types of methods would benefit from considering a Box-Cox transformation to be done on the data after the Fisher transformation has been applied.

Regarding the relationship between the mean and the variance of connectivity time series, we have previously discussed the mean-variance relationship in dynamic functional connectivity analysis and its implications for choice of analysis strategy (Thompson & Fransson 2015). In that study we explored the variance within a bounded range, i.e. without a Fisher transformation, which was performed because we, at that time, were concerned about the Fisher transformation for the reasons addressed in this paper. However, it was unclear from the Thompson and Fransson 2015 study whether the thresholding problems we addressed in that study persists for the case when the dFC time series variance was stabilized properly. To this end, we have in the present study explicitly shown when the variance has been stabilized (to the best of our ability) with first the Fisher, then the Box-Cox transformation, we are still able to replicate our previous findings related to the mean-variance dichotomy when defining thresholds of interesting dynamical activity in the dFC fMRI time series. The implications of this relationship might be substantial as it suggests that different analysis strategies (including both choices of thresholding and clustering techniques) may be biased by the variance differences among dFC time series. We suggest that scaling (for example by normalizing the data including demeaning and dividing the dFC time series by their standard deviation) should be performed to get all the dFC time series on equal mean and variance footing, which may be appropriate in many instances of data clustering and comparisons between conditions or group of subjects.

To conclude, the Fisher transform followed by a Box-Cox transform when applied on dFC fMRI data showed superior performance terms of stabilizing the dFC time series variance across its entire range. A possible drawback with the suggested approach is that the parameter λ needs to be fitted which poses additional work for the researcher. Perhaps, for efficiency, researchers could use a single λ value for all dFC time series (e.g. 1) but it would require that one is able to show that this single choice of λ has the desired effect in terms of stabilizing the variance. It is worth noting that when implementing the Box-Cox transform on economic data, Nelson & Granger (1979) often failed to find an adequate λ estimate for their model. Indeed, we occasionally observed that the λ parameter could at times fail to fit, but these instances were relatively few in number. Potentially, an expansion of the possible range of λ for these edges could be an option. Finally, due to the problem of negative values when λ=0, we scaled all time series to 1 and scaled them back to their original values after the transformation but other options exist such as fixing a non-zero λ to all time series.

# Acknowledgements

P.F. was supported by a grant from the Swedish Research Agency (grant no: 621-2012-4911).

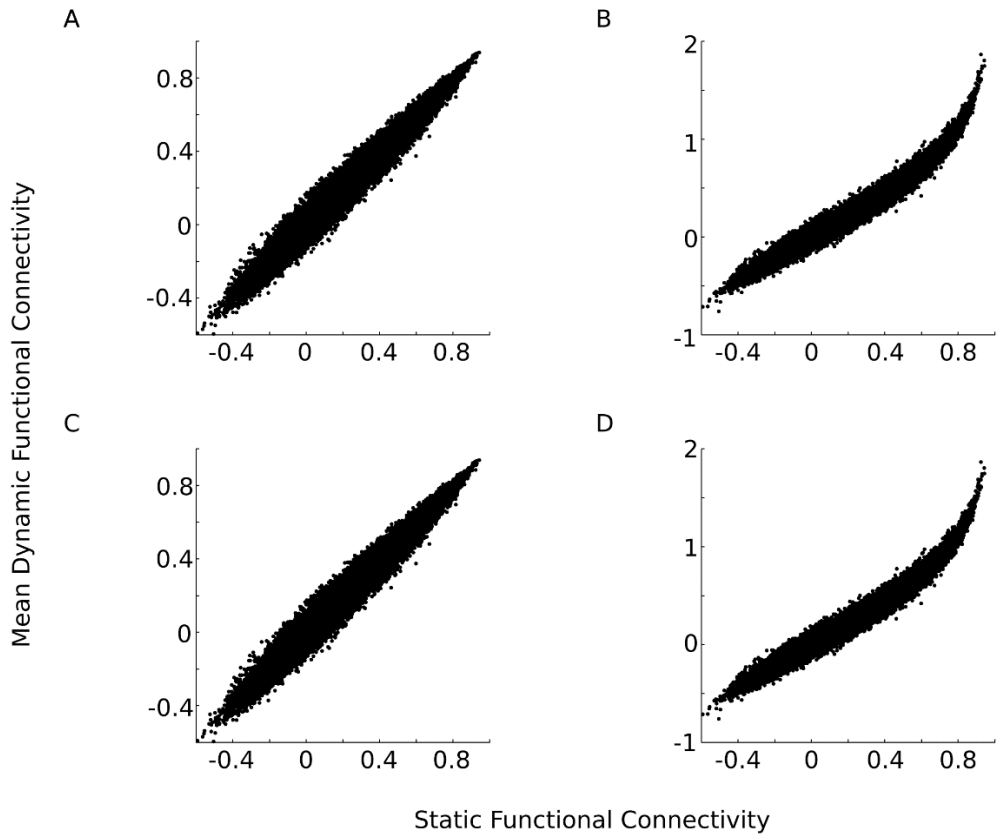

*Supplementary Figure 1*: Static functional connectivity shows a strong relationship with the mean of dynamic functional connectivity time series following different variance stabilizing transformations. Each point shows an edge in the default mode network for one subject (all DMN edges and all subjects are plotted). Each panel shows the static functional connectivity versus (A) the raw dynamic functional connectivity time series; (B) Fisher transformed connectivity values; (C) Box Cox transformed (D) Fisher followed by the Box-Cox transform.